\documentclass[aps,prc,twocolumn,groupedaddress,showpacs]{revtex4}

\usepackage{graphicx}

\begin{document}
%\preprint{preprint}

\title{Symmetry Energy, Temperature, Density and Isoscaling Parameter as a Function of Excitation energy
in A $\sim$ 100 mass region}
\author{D.V. Shetty, S.J. Yennello, G.A. Souliotis, A.L. Keksis, S.N. Soisson, B.C. Stein, and S. Wuenschel}
\affiliation{Cyclotron Institute, Texas A$\&$M University, College Station, Texas 77843, USA}
\date{\today}

\begin{abstract}
The symmetry energy, temperature, density and isoscaling parameter, in $^{58}$Ni + $^{58}$Ni, $^{58}$Fe + $^{58}$Ni 
and $^{58}$Fe + $^{58}$Fe reactions at beam energies of 30, 40 and 47 MeV/nucleon, are studied as a function of
excitation energy of the multifragmenting source. It is shown that the decrease in the isoscaling
parameter is related to the near flattening of the temperature in the caloric curve, and the 
decrease in the density and the symmetry energy with increasing excitation energy.  The decrease in 
the symmetry energy is mainly a consequence of decreasing density with increasing excitation rather than the increasing 
temperature. The symmetry energy as a function of density obtained from the correlation is in close agreement with 
the form,  E$_{sym}(\rho)$ $=$ 31.6 ($\rho/\rho_{\circ})^{0.69}$.
\end{abstract}


%\keywords{}
\maketitle

Due to its vast implications ranging from how nucleons clusterize into nuclei at low densities to the structure 
and stability of neutron stars at high density, the interest in understanding the behavior of nuclear matter at 
temperatures, densities and isospin (neutron-to-proton asymmetry) away from those of normal nuclear matter 
($T$ $\approx$ 0 MeV; $\rho_{o}$ $\approx$ 0.16 fm$^{-3}$; $N$ $\approx$ $Z$) has gained tremendous
importance \cite{BAL01}. 
\par
Experimentally, the multifragmentation reaction \cite{GRO90,BON95,BON85,FRI90,BOT02}, where a highly excited nucleus 
expands to a low density region and disassembles into many light and heavy fragments,  provides the best possible means of 
studying nuclear matter at non-normal densities, temperatures and isospin. It has been shown from 
various experiments \cite{NAT02} that the temperature as a function of excitation energy 
({\it {caloric curve}}) in multifragmentation reactions shows a near flattening, or a plateau-like region, at higher 
excitation energies. 
There are also indications that the density of the system decreases with increasing excitation energy \cite{NATO02,VIO04}. 
Recently, it has been shown that the isoscaling parameter, obtained from the fragment yield
distribution, shows a decrease with increasing beam energy \cite{SHE04,IGL06}. A decrease in the symmetry energy has also 
been experimentally observed \cite{SHE05,IGL06}.
\par
In this paper, we seek to understand the correlation between the temperature, density and symmetry energy
 of a multifragmenting system as it evolves with excitation energy. Such a correlation is important for constructing the nuclear matter 
equation of state and studying the density dependence of the symmetry energy; a key unknown in the
equation of state of asymmetric nuclear matter.
\par
We make use of the fragment yield distributions measured \cite{RAM98,SHE03,SHE04} in $^{58}$Ni, $^{58}$Fe + $^{58}$Ni, $^{58}$Fe 
reactions at 30, 40 and 47 MeV/nucleon to determine the isoscaling parameter $\alpha$, as a function 
of the excitation energy of the fragmenting source. The parameter $\alpha$'s were obtained by taking
the ratio of the isotopic yields for two different pairs of reactions, $^{58}$Fe + $^{58}$Ni and $^{58}$Ni +
$^{58}$Ni, and  $^{58}$Fe + $^{58}$Fe and $^{58}$Ni + $^{58}$Ni as described in Ref. \cite{SHE03,SHE04}. 
The excitation energy of the source for each beam energy was determined by simulating the initial stage of 
the collision dynamics using the Boltzmann-Nordheim-Vlasov (BNV) model calculation \cite{BAR02}.  The results 
were obtained at a time around 40 - 50 fm/c after the projectile had fused with the target nuclei and the quadrupole 
moment of the nucleon coordinates (used for identification of the deformation of the system) approached zero. These 
excitation energies were compared with those obtained from the systematic calorimetric measurements (see
Ref. \cite{NAT02}) for systems with mass ($A$ $\sim$ 100), and similar to those studied in the present work, and are 
in good agreement. Fig. 1 shows the experimental 
isoscaling parameter $\alpha$, as a function of the excitation energy obtained in the present study for both the pairs of 
reactions. A systematic decrease in the absolute values of the isoscaling parameter with increasing excitation energy is 
observed for both pairs. The $\alpha$ parameters for the $^{58}$Fe + $^{58}$Fe and $^{58}$Ni + $^{58}$Ni are about twice 
as large compared to those for the $^{58}$Fe + $^{58}$Ni and $^{58}$Ni + $^{58}$Ni pair of reactions. It is interesting 
to note that the difference in $Z/A$, {\it {i.e.}} $\Delta (Z/A)^{2}$, of the composite systems in the $^{58}$Fe + $^{58}$Fe and 
$^{58}$Ni + $^{58}$Ni pair is also twice as large compared to the $^{58}$Fe + $^{58}$Ni and $^{58}$Ni + $^{58}$Ni pair.   
%Fig 1
    \begin{figure}
    \includegraphics[width=0.5\textwidth,height=0.3\textheight]{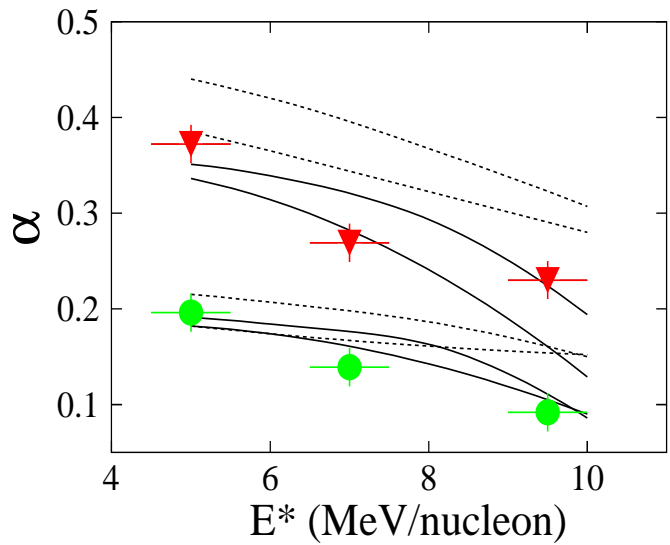} 
    \caption{Experimental isoscaling parameter $\alpha$, as a function of excitation energy for the Fe + Fe 
    and Ni + Ni pair of reaction (inverted triangles), and Fe + Ni and Ni + Ni pair of reactions (solid circles) for the 30, 
    40 and 47 MeV/nucleon. The solid and the dotted curves are the statistical multifragmentation model calculations as 
    discussed in the text.}
    \end{figure}
   
%Fig 2
    \begin{figure}
    \includegraphics[width=0.5\textwidth,height=0.3\textheight]{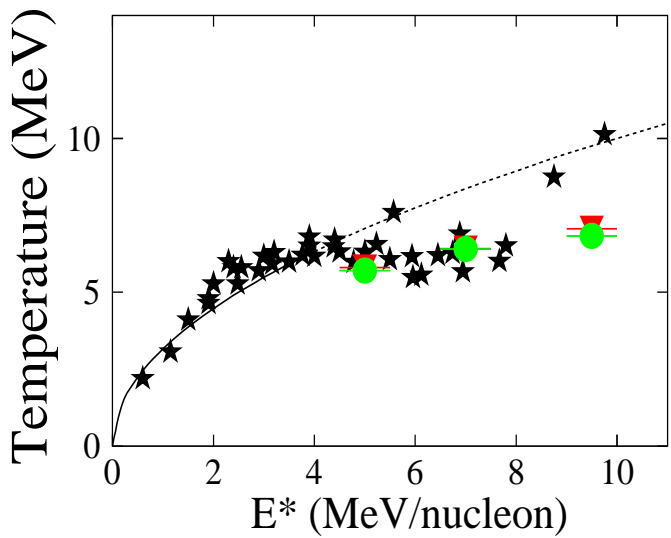} 
    \caption{Temperature as a function of excitation energy for the Fe + Fe and Ni + Ni pair of
    reaction (inverted triangles), and Fe + Ni and Ni + Ni pair of reactions (solid circles) for the 30, 40 and 47
    MeV/nucleon. The solid stars correspond to the measured values and are taken from Ref. \cite{NAT02}. The solid and the
    dotted curve corresponds to the Fermi-gas relation.}
    \end{figure}

\par
The experimental isoscaling parameter was compared to the predictions of the Statistical Multifragmentation 
Model (SMM) \cite{BON95,BOT01} calculations to study the dependence on the excitation energy and the
isospin content.  The initial parameters such as, the mass, charge and excitation energy of the fragmenting source 
for the calculation, was obtained from the BNV calculations as discussed above. To account for the possible uncertainties 
in the source parameters due to the loss of nucleons during pre-equilibrium emission, the calculations were
also performed for smaller source sizes. The break-up density in the calculation was taken to be multiplicity-dependent 
and was varied from approximately 1/2 to 1/3 the saturation density. This was achieved by varying the free volume with 
the excitation energy as shown in Ref. \cite{BON95}. The form of the dependence was adopted from the work of 
Bondorf {\it {et al.,}} \cite{BON85,BON98} (and shown by the solid curve in Fig. 4). It is known that the 
multiplicity-dependent break-up density, which corresponds to a fixed interfragment spacing and constant pressure 
at break-up, leads to a pronounced plateau in the caloric curve \cite{BON85,BON98}. A constant break-up density would 
lead to a steeper temperature versus excitation energy dependence. We will return to this point later in this paper.
\par
The symmetry energy in the calculation was varied until a reasonable agreement between the calculated and the 
measured $\alpha$ was obtained. It has been shown \cite{TSAN01,TSANG01,BOT02}, that the symmetry energy in the statistical model
calculations is related to the isoscaling parameter through the relation,

\begin{equation}
      \alpha ^{prim} = \frac{4C_{sym}}{T} {[(Z/A)_{1}^{2} - (Z/A)_{2}^{2}]}
\end{equation}

where $\alpha^{prim}$, is the isoscaling parameter for the hot primary fragments, i.e., before they
sequentially decay into  cold secondary fragments.  $Z_1$, $A_1$ and $Z_2$, $A_2$ are the charge and the 
mass numbers of the fragmenting systems. $T$ is the temperature of the systems and $C_{sym}$ is the symmetry energy.
In the above equation, the entropic contribution to the symmetry free energy is assumed to
be small (the contribution becomes important at densities below 0.008 $fm^{-3}$ \cite{HOR05}), the
symmetry energy can therefore be reliably substituted for the free energy.
\par
Fig. 1 shows the comparison between the SMM calculated and the measured $\alpha$ for both pairs of systems. The dotted 
curves correspond to the calculation for the primary fragments and the solid curves to the secondary
fragments. The width in the curve is the measure of the uncertainty in the inputs to the SMM calculation. It is observed 
that, within the given uncertainties,  the decrease in the $\alpha$ values with increasing excitation energy and 
decreasing isospin difference $\Delta(Z/A)^2$, of the systems is well reproduced by the SMM calculation. One also 
notes that the effect of sequential decay effect on the isoscaling parameter is small as has been observed in several 
other studies using statistical models \cite{TSAN01,TAN01}. 
\par
We show in Fig. 2, the temperature as a function of excitation energy ({\it {caloric curve}}) obtained from the above 
SMM calculation that uses the excitation energy dependence of the break-up density
to explain the observed isoscaling parameters. These are shown by the solid and inverted triangle
symbols. Also shown in the figure are the experimentally measured caloric curve data
compiled by Natowitz {\it {et al.}} \cite{NAT02} from various measurements for this mass range.  
The data from these measurements are shown collectively by solid
star symbols and no distinction is made among them. The Fermi-gas model predictions with inverse level
density parameter $K_{o}$ = 10 (solid and dashed lines), is also shown. It is evident from the 
figure that the temperatures obtained from the SMM calculations are in good agreement with the overall trend 
of the caloric curve. Somewhat lower value for the temperature is observed when the 
break-up density of the system is kept constant at 1/3 the normal nuclear density. By allowing the break-up density 
to evolve with the excitation energy, a near plateau that agrees with the experimentally measured caloric curves 
is obtained. This assures that the input parameters used in the SMM calculation for comparing with the data are
reasonable.

%Fig 3
    \begin{figure}
    \includegraphics[width=0.5\textwidth,height=0.3\textheight]{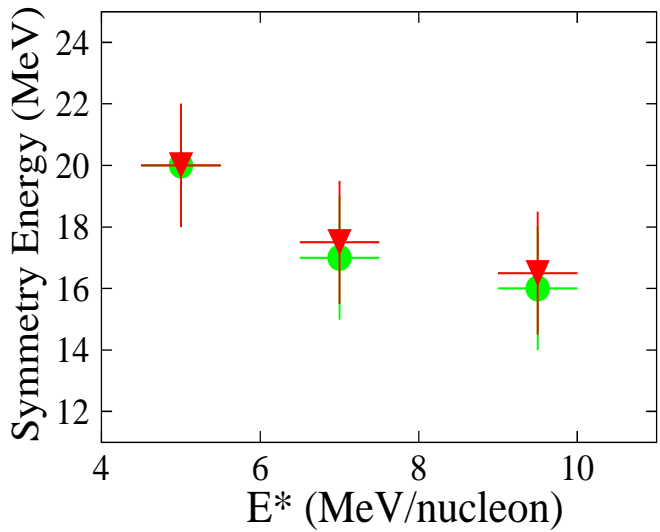} 
    \caption{Symmetry energy as a function of excitation energy for the Fe + Fe and Ni + Ni pair of
    reactions (inverted triangles), and Fe + Ni and Ni + Ni pair of reactions (solid circles) for the 30, 40 and 47
    MeV/nucleon.}
    \end{figure}

%Fig 4
    \begin{figure}
    \includegraphics[width=0.5\textwidth,height=0.30\textheight]{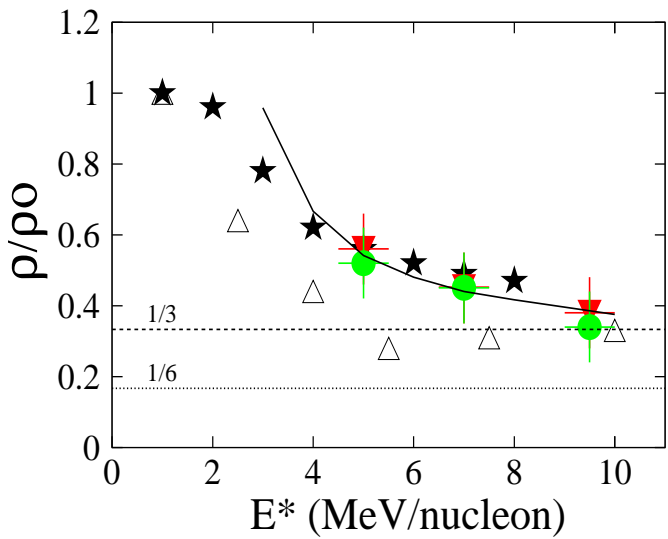} 
    \caption{Density as a function of excitation energy for the Fe + Fe and Ni + Ni pair of
    reactions (inverted triangles), and Fe + Ni and Ni + Ni pair of reactions (solid circles) for the 30, 40 and 47
    MeV/nucleon. The solid stars correspond to those taken from Ref. \cite{NATO02}. The open triangles are those
    from Ref. \cite{VIO04}. The solid curve is from Ref. \cite{BON85}.}
    \end{figure}

\par
The symmetry energies obtained from the statistical model comparison of the experimental isoscaling
parameter $\alpha$, are as shown in Fig. 3. A steady decrease in the symmetry energy with increasing excitation 
energy is observed for both pairs of systems. Such a decrease has also been observed in several other 
studies \cite{IGL06,SHE05,FEV05,HEN05}. We have also estimated the effect of the
symmetry energy evolving during the sequential de-excitation of the primary fragments \cite{BUY05,IGL06}. These are reflected in
the large error bars shown in Fig. 3.
\par
The phase diagram of the multifragmenting system is two dimensional and hence the excitation energy
dependence of the temperature (the caloric curve) must take into account the density dependence too.
Often this dependence is neglected while studying the caloric curve. In the following, we attempt to extract 
the density of the fragmenting system as a function of excitation energy. It has been shown by 
Sobotka {\it {et al.}} \cite{SOB04}, that the plateau in the caloric curve could be a consequence of 
the thermal expansion of the system at higher excitation energy and decreasing density. By assuming that the 
decrease in the breakup density, as taken in the present statistical multifragmentation calculation, can be 
approximated by the expanding Fermi gas model, and furthermore the temperature in Eq. 1 and the temperature in the 
Fermi-gas relation are related, one can extract the density as a function of excitation energy using the 
relation

\begin{equation}
      T =  \sqrt {K(\rho) E^*} = \sqrt {K_{o} (\rho / \rho_o)^{2/3} E^*}
\end{equation}

In the above expression, the momentum and the frequency dependent factors in the effective mass ratio are 
taken to be one as expected at high excitation energies and low densities studied in this work
\cite{HAS86,SHL90,SHL91}.
\par
The resulting densities for the two pairs of systems are shown in Fig. 4 by
the solid circles and inverted triangles. For comparison, we also show the break-up densities obtained from  
the analysis of the apparent level density parameters required to fit the measure caloric curve by
Natowitz {\it {et al.}} \cite{NATO02} and those obtained by Viola {\it {et al.}} \cite{VIO04} from the Coulomb 
barrier systematics that are required to fit the measured intermediate mass fragment kinetic energy spectra. One 
observes that the present results obtained by requiring to fit the measured isoscaling parameters and the caloric 
curve are in good agreement with those obtained by Natowitz {\it {et al}}. The figure also shows the fixed 
freeze-out density of 1/3 (dashed line) and 1/6 (dotted line) the saturation density assumed in various 
statistical model comparisons. 
\par
It is evident from figures 1, 2, 3 and 4 that the decrease in the experimental isoscaling parameter $\alpha$, 
symmetry energy, break-up density, and the flattening of the temperature with increasing excitation energy 
are all correlated. One can thus conclude that the expansion of the system during the multifragmentation process 
leads to a decrease in the isoscaling parameter, decrease in the symmetry energy and density, and the 
flattening of the temperature with excitation energy. Table I shows the correlated quantities 
obtained using Eqs. 1 and 2 for both pairs of systems.
\par
%Fig 5
    \begin{figure}
    \includegraphics[width=0.5\textwidth,height=0.30\textheight]{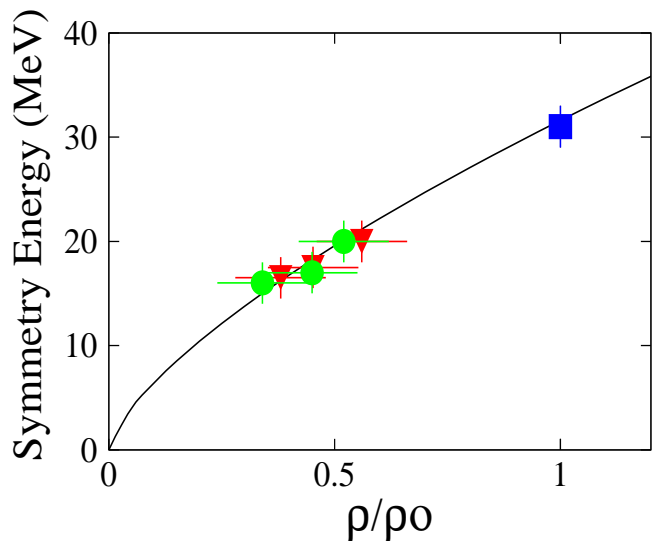} 
    \caption{Symmetry energy as a function of density for the Fe + Fe and Ni + Ni pair of
    reaction (inverted triangles), and Fe + Ni and Ni + Ni pair of reactions (solid circles) for the 30, 40 and 47
    MeV/nucleon. The solid square corresponds to those from Ref. \cite{KHO05}. The solid curve is the
    dependence obtained in Ref. \cite{SHET05,SHET06}.} 
    \end{figure}

\begin{table*}
\caption{\label{tab:table1} Primary isoscaling parameter, temperature, density and
symmetry energy obtained for Fe + Fe and Ni + Ni pair (top row), and Fe + Ni and Ni + Ni pair (bottom
row) of systems at various excitation energies using Eqs. 1 and 2.}
\begin{ruledtabular}
\begin{tabular}{cccccccc}
 E$^{*}$ (MeV/nucleon) & $\alpha ^{prim}$         &  Temp. (MeV)      &  $\rho$/$\rho_o$  & Symmetry Energy (MeV)  \\
\hline
 5.0                   &  0.44                    &  5.8              &  0.56             &  20.0                  \\
 7.0                   &  0.35                    &  6.4              &  0.45             &  17.5                  \\
 9.5                   &  0.30                    &  7.0              &  0.38             &  16.5                  \\  
\hline
 5.0                   &  0.22                    &  5.7              &  0.52             &  20.0                  \\
 7.0                   &  0.17                    &  6.4              &  0.45             &  17.0                  \\
 9.5                   &  0.15                    &  6.8              &  0.34             &  16.0                  \\  
\end{tabular}
\end{ruledtabular}
\end{table*}

From the above correlation between the symmetry energy as a function of excitation energy, and the density as a
function of excitation energy, we obtain the symmetry energy as a function of density. This is shown in Fig. 5. 
for both pairs of systems. The temperature in the present work remains nearly constant for the range of excitation energies studied,
the  observed decrease in the symmetry energy with increasing excitation energy is therefore a
consequence of decreasing density. This is also supported by microscopic calculations which shows an
extremely slow evolution of the symmetry energy with temperature \cite{BAL01,BAL06}. The evolution is practically negligible 
for the temperature range studied in this work. Also shown in Fig. 5 is the symmetry energy
value of Khoa {\it {et al.}} \cite{KHO05}, (solid square) obtained by fitting the experimental differential cross-section data in a
charge exchange reaction using the isospin dependent optical potential. The 
solid curve corresponds to the dependence, E$_{sym} (\rho)$ $=$ 31.6 ($\rho/\rho_{\circ})^{0.69}$, obtained by comparing the present data with 
the Antisymmetrized Molecular Dynamic (AMD) calculation
\cite{ONO03}, in previous work \cite{SHE04,SHET05,SHET06}. The similarities in the density 
dependence of the symmetry energy obtained from the present statistical model approach and the AMD model approach
is intriguing. It should be noted that the symmetry energy shown by the solid curve in the figure relates to
the volume part of the symmetry energy as in
infinite nuclear matter, whereas, the symmetry energy obtained from the present study (solid
circles and inverted triangles) relates to the fragment that are finite and has surface contribution. 
The similarity between the two can be understood in terms of the weakening of the surface symmetry free energy when the fragments
are being formed. During the density fluctuation in uniform low density matter, the fragments are not completely 
isolated and continue to interact with each other, resulting in a decrease in the surface contribution as predicted by Ono {\it {et al.}}
\cite{ONO04}. The present observation therefore lends credence to the fact that it is possible to directly obtain the properties of infinite nuclear matter from the  fragments produced in the 
multifragmentation process \cite{ONO04}. More studies are required to illustrate this point further.
\par
In summary, we have studied the isoscaling parameter, symmetry energy, temperature and density as a function of 
excitation energy in reactions populating A $\sim$ 100 mass region. It is observed that the decrease in the experimental 
isoscaling parameter $\alpha$, symmetry energy, breakup density, and the flattening of the temperature with
increasing excitation energy are related to each other. The  observed decrease in the symmetry energy with 
increasing excitation energy appears to be mainly a consequence of decreasing density. The symmetry energy
as a function of density obtained from the present study is consistent with those obtained from the
dynamical AMD calculation, indicating that the surface contribution to the symmetry energy could be small. 
 
\section{Acknowledgment}
We thank Drs. J.B. Natowitz and V. Viola for providing us the data points from their studies on break-up
density and caloric curve. We also thank A. Botvina for fruitful discussion and the Catania group for
their BNV code. This work was supported in part by the Robert A. Welch Foundation through grant No. A-1266, 
and the Department of Energy through grant No. DE-FG03-93ER40773.

\end{document}